\documentclass[preprint,prl,aps,amsmath,amssymb,color,superscriptaddress]{revtex4}
\usepackage{stmaryrd}
\usepackage{amsmath}
\usepackage{amssymb}
\usepackage{graphicx}
\usepackage{dcolumn}
\usepackage{bm}
\usepackage{tabularx}
\begin{document}
\begin{titlepage}
\title{Robust high-temperature topological excitonic insulator of transition-metal carbides (MXenes)}
\author{Shan Dong}
\affiliation{Key Lab of advanced optoelectronic quantum architecture and measurement (MOE), and Advanced Research Institute of Multidisciplinary Science, Beijing Institute of Technology, Beijing 100081, China}
\author{Yuanchang Li}
\email{yuancli@bit.edu.cn}
\affiliation{Key Lab of advanced optoelectronic quantum architecture and measurement (MOE), and Advanced Research Institute of Multidisciplinary Science, Beijing Institute of Technology, Beijing 100081, China}
\date{\today}

\begin{abstract}
Topological excitonic insulators combine topological edge states and spontaneous exciton condensation, with dual functionality of topological insulators and excitonic insulators. Yet, they are very rare and little is known about their formation. In this work, we find that a mechanism dubbed as parity frustration prevents excitonic instability in usual topological insulators, and those whose band inversion is independent of spin-orbit coupling are possible candidates. We verify this by first-principles calculations on monolayer transition-metal carbides (MXenes), which show a robust thermal-equilibrium exciton condensation, being sufficient for topological applications at room temperature. Such a state can be identified by angle-resolved photoemission spectroscopy and transport measurement. Our work provides not only a guide for finding more topological excitonic insulators, but also a new platform for studying the interplay between non-trivial band topology and quantum many-body effects.
\end{abstract}

\maketitle
\draft
\vspace{2mm}
\end{titlepage}

Due to the interplay between topological order and symmetry-breaking, topological materials with coexisting quantum many-body phase bring new opportunities for exploring many novel physical phenomena, such as the quantum anomalous Hall effect and Majorana fermions when coupled with magnetism and superconductivity, respectively\cite{Rienks,YuR,topoS}. Topological excitonic insulator (TEI) is a new topological state of matter that combines non-trivial band topology and spontaneous exciton condensation\cite{Budich,Pikulin,Du,Yu,Varsano,Jia,Sun}. Therein topological edge states exist inside a bulk gap by the condensation of spontaneously formed excitons (Coulomb bound electron-hole pairs), as the case in excitonic insulators\cite{Mott,Kohn,Halperin}, rather than by the spin-orbit coupling (SOC) in usual topological insulators. In this sense, the TEI can be viewed as a superposition of topological insulators and excitonic insulators as shematically illustrated in Fig. 1. The excitonic insulator is a macroscopic quantum system with a reconstructed many-body ground state akin to the superconductor. It forms when the exciton binding energy ($E_b$) exceeds the one-electron gap ($E_g$), accompanied by symmetry-breaking\cite{VarsanoNC2017,Jiang2019,Kaneko,Mazza}. So far, however, no material has been universally accepted as an excitonic insulator.

The TEI bridges topological insulator and excitonic insulator, and thus endows an innate advantage to address some key issues in the two fields. On the topological side, despite many efforts, the topological insulator that can support room temperature applications is still in pursuit to materialize\cite{Rienks}. Whilst, for the TEI, the gap is renormalized to a larger value by the exciton condensation, becoming not proportional to the SOC strength. This not only helps to raise the working temperature of relevant devices but also makes it possible to design large-gap topological materials comprised of only relatively light elements. Furthermore, due to unavoidable defects, bulk carriers are ubiquitous in conventional SOC topological insulators, which can mask or drown out signals from the topological boundary states. In practice, carrier compensation is often used to tune the Fermi energy into the bulk gap. Instead, the bulk gap of the TEI originates from the preformed excitons associated with the overall screening characteristics of the system, so the effect of carriers introduced by defects would be greatly suppressed.

On the excitonic side, verification of spontaneous exciton condensation remains an insurmountable challenge\cite{Chang,Liu,Kogar}, partly because charge-neutral nature of excitons prevents the detection of their coherent flow by means of electrical transport measurements as in the case of superconductivity. Now, experimental physicists could look for clues in transport signatures given the emergence of conductive edge states in the TEIs. Meanwhile, the TEI may exhibit some unique properties beyond both\cite{Hu,Wang}. One of its remarkable features is the complete super-transport combining bulk excitonic superfluidity and dissipationless edge currents, which could provide a broad and unprecedented application potential in electronics and spintronics. While the superfluidity caused by exciton condensation is still under debate\cite{Halperin,Sham,Mazza}, it is certain that exciton flow can carry information and energy. Then, the TEI allows people to integrate the two cornerstones of today's information technology into a same material through using edge states for signal processing and using bulk excitons for signal communication\cite{Jiang2019}.

Nevertheless, despite its many merits, the TEI is still in infancy, especially limited by material realizations. To date, the hints are only reported in InAs/GaSb quantum wells\cite{Pikulin,Du,Yu} and transition-metal dichalcogenides\cite{Varsano,Jia,Sun}. These candidates have too small many-body gaps, making it practically difficult to distinguish excitonic instability from other possible competing mechanisms\cite{Kogar,Du,Yu,Varsano,Jia,Sun} and limiting future devices to operate at low temperatures. There is also no understanding about what kinds of materials make TEIs, which is essential for finding and predicting more TEIs.

In this work, we first present an explanation, dubbed as parity frustration, for the rarity of the TEI. Based on this analysis, we point out that usual topological insulators with band inversion independent of the SOC provide fertile ground for realizing such kind of quantum materials. We then demonstrate this with first-principles calculations on monolayer transition-metal carbides (MXenes). We find a robust thermal-equilibrium exciton Bose-Einstein condensation (BEC) against in-plane strain and vertical electric-field, even with the critical temperature up to room temperature and above, as well. We will discuss different physics due to the unique coupling of band topology and exciton BEC in these MXenes, as well as the specific signature that permits the experimental identification of the TEI. Finally, we briefly introduce the recent progress on the synthesis of such MXenes.

All density functional theory (DFT) calculations were performed within the Perdew-Burke-Ernzerhof (PBE) generalized gradient approximation\cite{Perdew} as implemented in the Quantum ESPRESSO package\cite{Giannozzi}. Fully relativistic norm-conserving Vanderbilt pseudopotentials\cite{Hamann,Chiang} were employed with an energy cutoff of 55 Ry. A vacuum layer of more than 12 \AA\ was applied along the out-of-plane direction in order to minimize spurious interactions with its replica. An 18 $\times$ 18 $\times$ 1 $\Gamma$-centered $k$-point grid was used to sample the Brillouin zone. The lattice constants and atomic positions were fully relaxed until residual force on each atom was less than 0.01 eV/\AA. The many-body perturbation theory within the $G_{0}W_{0}$ approximation\cite{Hybertsen} was employed using the YAMBO code\cite{Marini} to cure the DFT bandgap problem. The low-energy excitonic properties were calculated by solving the Bethe-Salpeter equation (BSE)\cite{Rohlfing} with the Coulomb cutoff technique. In solving the BSE, the scissor operator, which is taken as the gap correction from PBE to $G_{0}W_{0}$ at the $\Gamma$ point (see Table S1\cite{SI}), is used to correct the quasi-particle energies and introduced both for the response function and diagonal part of the BSE kernel\cite{Sangalli}. Top four valence bands and bottom four conduction bands are chosen to build the BSE Hamiltonian. A fine 33 $\times$ 33 $\times$ 1 $k$-point grid, 260 bands, and 12 Ry cutoff were used to evaluate the dielectric function matrix.

\begin{figure}[htbp]
\includegraphics[width=0.85\columnwidth]{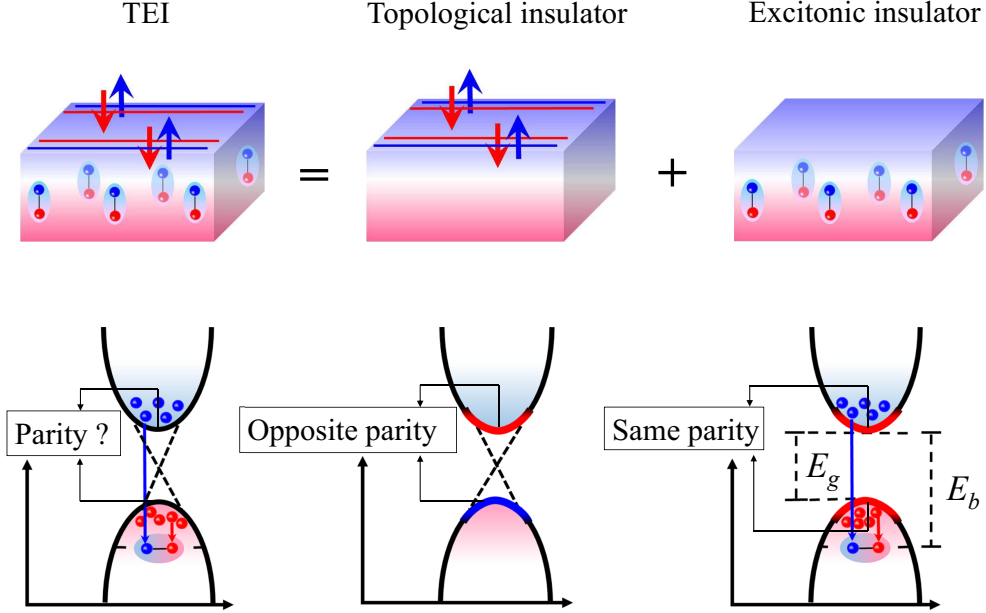}
\caption{\label{fig:fig1} Schematic of a TEI as a superposition of a topological insulator and an excitonic insulator. Blue and red balls are electrons and holes, respectively, and shadowed ellipses indicate their strong binding as excitons. $E_{g}$ is the one-electron gap while $E_{b}$ is the exciton binding energy.}
\end{figure}

\begin{figure*}[htbp]
\includegraphics[width=0.95\columnwidth]{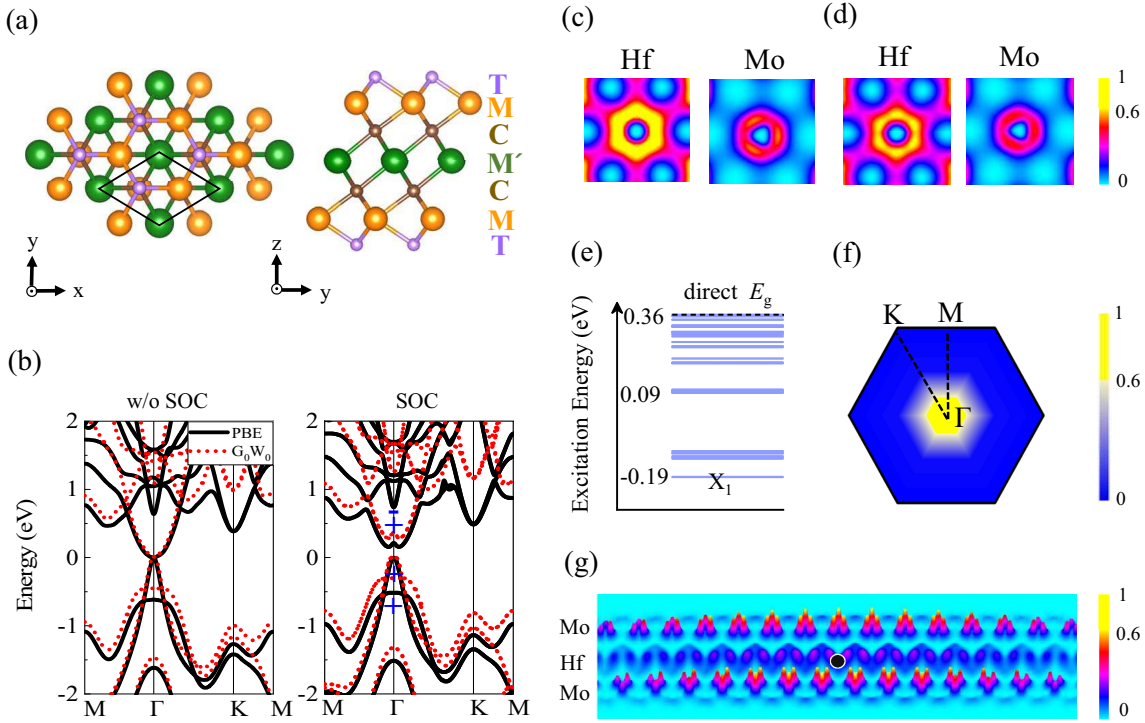}
\caption{\label{fig:fig2} Structural, electronic and excitonic properties of monolayer M$ _{2} $M$ ^{'} $C$ _{2} $T$ _{2} $ MXenes. (a) Crystal structure of M$ _{2} $M$ ^{'} $C$ _{2} $T$ _{2} $. M, M$ ^{'} $ and T stand for (Mo, V), (Hf, Zr, Ti) and (O, F), respectively. Black rhombus denotes the unit cell. (b) Band structures of Mo$ _{2} $HfC$ _{2} $O$ _{2} $ without and with the SOC. ``+'' and ``-'' denote even and odd parity, respectively. Cross sections of decomposed charge densities at the $\Gamma$ point for the (c) top valence band and (d) bottom conduction band through Hf and Mo atomic planes. (e) Excitation energy spectrum for $ q = 0 $ excitons. Each horizontal line represents an exciton state. (f) Reciprocal-space wavefunctions modulus for the $X_{1}$-exciton. (g) Two-dimensional cross section in $ yz $ plane of real-space wavefunctions for the $ X_{1}$-exciton, which contains 11 unit cells. Black dot denotes the hole position. For clarity, the maximum value of charge densities/wavefunctions modulus has been renormalized to unity and truncated at 60$\%$ in (c), (d), (f) and (g).}
\end{figure*}

If an excitonic instability occurs in a conventional SOC topological insulator, it leads straightforwardly to the TEI. However, as illustrated in Fig. 1, this is inhibited by what we call the parity frustration. Without losing generality, let us consider a gapped system with inversion symmetry. To be topologically non-trivial, there is often parity inversion between frontier bands\cite{Fu}, which requires these relevant states to have opposite parities. In contrast, occurrence of the excitonic instability prefers the band-edge states with the same parity to break the synchronous change between $E_b$ and $E_g$, and eventually to realize $E_b > E_g$\cite{Halperin,Duan,Jiang,DongAlSb}. Thereby, the TEI encounters some parity frustration.

Above simple analysis gives an intuitive understanding of why the TEI is so rare, but in turn, it also points out the direction to search for the TEIs. It has become clear that different energy states have to be independently responsible for the non-trivial topology and the excitonic instability, so as to avoid the parity-frustration. Given that the topological insulators require the SOC to open a gap, in order to fulfill  the same parity for the frontier states across the SOC gap, the parity inversion has to be associated with other states rather than completely with the band-edge states. The $a$ $priori$ demand reminds us of those topological insulators with band inversion independent of the SOC, because they can provide the unique ``functional segregation", i.e., different states contribute to non-trivial topology and excitonic instability, respectively. Below, we illustrate how this ``functional segregation" comes into play to result in the TEIs using four double transition-metal MXenes M$ _{2} $M$ ^{'} $C$ _{2} $T$ _{2} $ (V$ _{2} $TiC$ _{2} $F$ _{2} $, Mo$ _{2} $TiC$ _{2} $O$ _{2}  $, Mo$ _{2} $ZrC$ _{2} $O$ _{2} $ and Mo$ _{2} $HfC$ _{2} $O$ _{2} $) as concrete examples.

These MXenes include seven atomic layers with an out-of-plane ordering and form a hexagonal lattice in each basal plane [see Fig. 2(a)]. Figure 2(b) typically shows for the Mo$ _{2} $HfC$ _{2} $O$ _{2} $ one-electron band structure, and the results for other systems are shown in Fig. S1\cite{SI}. Without the SOC, the band is gapless while after including the SOC, an $ E_{g} $ of 0.22/0.36 eV is opened at the $\Gamma$ point by the DFT/GW. Because the system possesses the inversion symmetry, we perform parity analysis\cite{Fu} and find that the Mo$ _{2} $HfC$ _{2} $O$ _{2} $ is a topological insulator due to $  Z _{2 }$  = 1, in line with previous studies\cite{Si,Khazaei,Huang}. However, unlike usual cases, here parity inversion does not occur between top valence band and bottom conduction band. Instead, both two frontier states have the same even parity and are dominantly contributed by the Hf 5$d$-orbitals [see Figs. 2(c) and 2(d)]. The same parity means dipole forbidden transitions in between, which will significantly reduce the electron-hole screening interaction and enhance the exciton binding. More importantly, it breaks the scaling university between $E_b$ and $E_g$ for two-dimensional semiconductors\cite{Duan} and probably trigger an excitonic instability\cite{Jiang}. Band nesting around the $\Gamma$ point also implies a strong exciton binding. So, let us first consider the excitonic property before discussing the non-trivial topology in more detail.

A negative excitation energy (defined as $E_t = E_g - E_b$) of -0.19 eV is found for the ground-state $X_1$-excitons in the Mo$ _{2} $HfC$ _{2} $O$ _{2} $, as shown in Fig. 2(e). This implies their spontaneous formation and the tendency to drive a transition into the exciton BEC. While no BEC exists for free bosons  in two dimensions, the situation is changed for systems with localized states below the continuum of extended states\cite{Jan}. This is exactly the case in our work. As we shall see later, the spontaneous production of excitons breaks the spatial symmetry. Not changing the time-reversal symmetry allows the non-trivial edge states to be maintained. Combined with its non-trivial topology, the occurrence of excitonic instability would lead monolayer Mo$ _{2} $HfC$ _{2} $O$ _{2} $ to become a TEI. As expected, the $X_1$-exciton is optically dark. Figure 2(f) shows for the $X_1$-exciton the reciprocal-space wavefunction. Its relatively localized distribution at the Brillouin-zone center corresponds to a large delocalization in the real-space, as manifested by a snapshot in the $yz$ plane [see Fig. 2(g)].

The formation of $X_1$-excitons causes Mo atoms that are all equivalent under the one-electron picture to become dynamically different, hence breaking the space-inversion symmetry. As reflected by Fig. 2(g), the presence of excitons makes the translation period no longer the unit cell size of the crystal (The exact value depends on factors such as concentration). But note that the exciton is mobile, i.e. the position of the hole in Fig. 2(g) can migrate from one lattice to another\cite{Jiang2019}. If the time scale considered allows the hole to travel through all lattice points, the spatial translation period on this time scale is still the unit cell size. However, it is clear that the translation periodicity is different in the two cases. Thus, due to the preformed excitons, the TEI phase would exhibit a completely different space-inversion symmetry, similar to that found in carbon nanotubes\cite{VarsanoNC2017}, transition-metal halides\cite{Jiang2019} and two-orbital model studies\cite{Kaneko}. This symmetry-breaking is entirely electronically driven and can occur in the absence of a change in the lattice.

The situation is fundamentally different from that in Ta$_2$NiSe$_5$ or TiSe$_2$\cite{Watson,Windg,Volkov,Mazza,Kogar,Kim}, which is accompanied by lattice distortions. In particular, whether the charge-density-wave transition in TiSe$_2$ comes from the Jahn-Teller mechanism or from the excitonic instability has plagued the excitonic insulator community for decades and remains unresolved. It is well known that the phonon spectrum of TiSe$_2$ has imaginary frequencies\cite{Calandra} and a supercell structure-optimization can yield the CDW superlattice structure\cite{Bianco}, called the Jahn-Teller mechanism. Excitonic instability is another hypothesis to explain the charge-density-wave phase transition therein. Which one it is has been debated for decades and is still ongoing. However, both our and previous calculations\cite{Si} show that there is no imaginary frequency on the phonon spectrum of MXenes, and even if we start from a structure with some degree of distortion, the relaxation still yields a symmetric structure. Experimentally, no structural distortions were found in these synthesized MXenes either\cite{Anasori,Meshkian}. Therefore, these MXenes inherently exclude Jahn-Teller-like instabilities, consequently avoiding the confusion of TiSe$_2$ and removes a daunting obstacle to the experimental confirmation that MXenes are excitonic insulators. But the lack of a ``structural" signal poses new challenges for identifying phase transitions and determining the nature of the gap\cite{Liu}. Although the flattening of the band-edge state can be used as a probe\cite{Cercellier}, as be shown later, the distinct SOC-dependence also plays a role for this purpose to distinguish whether the bulk gap originates from the spontaneous exciton condensation or the SOC.

\begin{figure}[htbp]
\includegraphics[width=0.8\columnwidth]{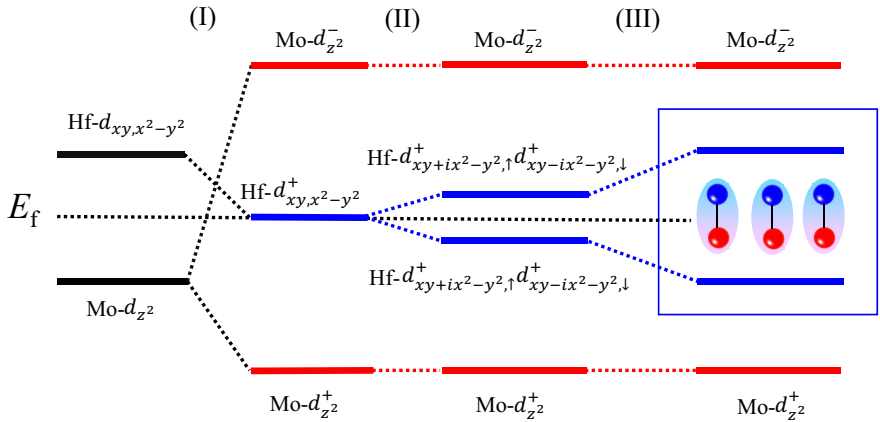}
\caption{\label{fig:fig3} Schematic from the atomic 4d/5d orbitals of Mo/Hf into the exciton BEC of Mo$ _{2} $HfC$ _{2} $O$ _{2} $ at the $\Gamma$ point. The stages (I), (II) and (\uppercase\expandafter{\romannumeral3}) represent the effect of turning on chemical bonding, SOC and electron-hole interaction. In the blue box, preformed excitons (shadowed ellipses) make the MXenes to be TEIs.}
\end{figure}

To get a physical understanding of its TEI nature, we consider the band evolution of the Mo$ _{2} $HfC$ _{2} $O$ _{2} $ at the $\Gamma$ point, which is schematically illustrated in three stages (\uppercase\expandafter{\romannumeral1}), (\uppercase\expandafter{\romannumeral2}) and (\uppercase\expandafter{\romannumeral3}) in Fig. 3. Upon crystal formation in stage (\uppercase\expandafter{\romannumeral1}), the 4$ d_{z^{2}} $ orbitals of two Mo atoms form bonding and antibonding states with opposite parities\cite{Si}. Now band topology has changed without the SOC, occurring between the reconstructed Mo antibonding state and the Hf 5$ d $ state. In stage (\uppercase\expandafter{\romannumeral2}), turning on the SOC lifts the double degeneracy at the Fermi energy, giving rise to a topologically non-trivial gap. Unlike most topological insulators where the SOC both opens the gap and causes the band inversion, here the SOC plays a single role, independent of band inversion. Consequently, the band-edge states are allowed to have the same parity. When the electron-hole interaction is taken into account in stage (\uppercase\expandafter{\romannumeral3}), the spontaneous $X_1$-exciton condensation is facilitated, leading to a many-body ground state responsible for the bulk insulation.

\begin{figure*}[htbp]
\includegraphics[width=0.95\columnwidth]{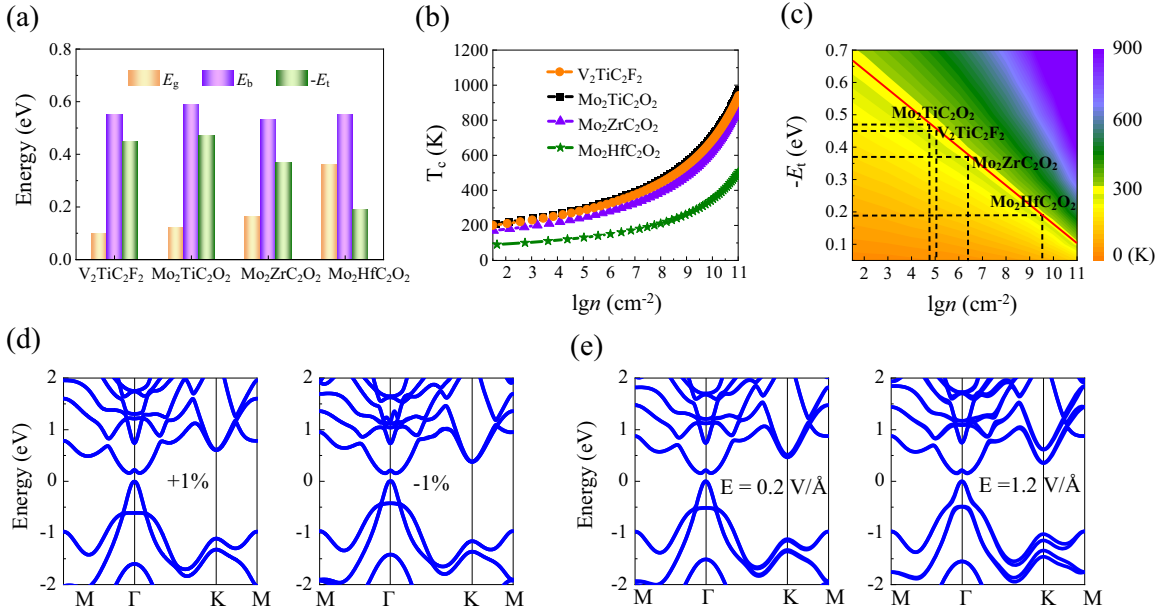}
\caption{\label{fig:fig4} (a) One-electron gap ($ E_{g} $), exciton binding energy ($ E_{b} $) and excitation energy ($ E_{t} $) of all M$ _{2} $M$ ^{'} $C$ _{2} $T$ _{2} $ MXenes. (b) Transition temperature $ T_{c} $ as a function of the log of exciton density $ n $. (c) Map of $ T_{c} $ as functions of ($\lg n$, -$ E_{t} $) with the exciton mass fixed to that of free-electron. The red line represents $ T_{c} = 300 $ K. Band structures of the Mo$ _{2} $HfC$ _{2} $O$ _{2 } $ under (d)  $\pm$1$ \% $ in-plane strain, and (e) vertical electric-field of 0.2 and 1.2 V/\AA .}
\end{figure*}

Similar physics exists for V$ _{2} $TiC$ _{2} $F$ _{2} $, Mo$ _{2} $TiC$ _{2} $O$ _{2}  $ and Mo$ _{2} $ZrC$ _{2} $O$ _{2} $. In Fig. 4(a), we summarize $  E_{g} $, $  E_{b} $ and $  E_{t} $ for all four MXenes, and the detailed results can be found in Table S1\cite{SI}. From Ti to Zr to Hf, the atomic SOC becomes stronger and stronger, so the $  E_{g} $ of corresponding compound gets larger and larger. Whilst, the  $  E_{b} $  is all around 0.55 eV, being almost system independent. This is very interesting because for the TEIs, their bulk gap is roughly measured by the $  E_{b} $\cite{Chang} rather than the SOC opened $E_{g} $. From this perspective, we estimate that the exciton BEC increases the gap of Mo$ _{2} $HfC$ _{2} $O$ _{2} $  by 53\%. This enhancement becomes more significant in the compound with weaker atomic SOC, e.g., as high as 4.5 times for the V$ _{2} $TiC$ _{2} $F$ _{2} $. So, the TEIs have an inherent advantage of being large-gap topological insulators but fully composed of light elements. On the other hand, the feature that the gap becomes disproportionate to the SOC would offer an observable evidence for identifying a spontaneous exciton BEC. This can be experimentally tested by angle-resolved photoemission spectroscopy, or by transport measurements in terms of topological edge modes.

The $E_{t} $, however, displays a distinctly different trend. For instance, the Mo$ _{2} $TiC$ _{2} $O$ _{2} $ has the second smallest $  E_{g} $ but the most negative $  E_{t} $. For usual topological insulators, the larger the SOC gap is, the higher the operating temperature is. Whereas for the TEIs, the situation is changed, as its bulk insulation arises from the exciton BEC whose critical temperature $T_{c}$ depends on the $E_{t}$ under two-dimensional cases\cite{ Jan}.

The $ T_{c}  $ is calculated by exciton density $ n $, exciton mass $m$ and the $  E_{t} $ using the formula\cite{ Jan}:
\[n = - \frac{1}{{\lambda {{_c^2}_{}}}}\ln (1 - {e^{ - {\beta _c}|{E_t}|}})\]
with $\lambda _c = (2\pi \hbar^{2}/mk_{B}T_{c})^{\frac{1}{2}}$ and ${\beta _c} = {({k_B}{T_c})^{ - 1}}$. Figure 4(b) presents obtained $ T_{c}  $ as a function of $ n $ for each MXene. All of the $ T_{c}  $ is very high. At $ n $ as low as 10$ ^{2} $ cm$ ^{-2} $, the lowest $ T_{c} $ reaches $\sim$100 K. With increase of the $ n $, the $ T_{c}  $ increases further. Under the same $ n $, the trend of $ T_{c}  $ is  Mo$ _{2} $TiC$ _{2} $O$ _{2} > $ V$ _{2} $TiC$ _{2} $F$ _{2} > $ Mo$ _{2} $ZrC$ _{2} $O$ _{2} > $ Mo$ _{2} $HfC$ _{2} $O$ _{2 } $, which is the same as -$  E_{t} $ [see Fig. 4(a)]. As revealed in Table S1, the $ m $ displays a distinct trend of Mo$ _{2} $TiC$ _{2} $O$ _{2} > $ V$ _{2} $TiC$ _{2} $F$ _{2} > $ Mo$ _{2} $HfC$ _{2} $O$ _{2 } > $ Mo$ _{2} $ZrC$ _{2} $O$ _{2} $ and varies in the range of 0.8 $\sim$ 1.6 $ m_{0} $ ($ m_{0} $ is the mass of free electron). Actually, we find that the $ T_{c}  $ is almost entirely determined by -$  E_{t} $ and the $ m $ shows a marginal effect only at very large $ n $. Without losing generality, we fix $  m = m_{0} $ and plot the $ T_{c}  $ dependences on ($ n $, -$  E_{t} $) in Fig. 4(c), whereby we derive a rough condition between $E_t$ and $n$ for room temperature TEIs, namely, $  E_{t}  = 0.06 \lg n - 0.76$. For here studied MXenes, this corresponds to a density range of 10$ ^{5}$ $\sim$ 10$ ^{10} $ cm$ ^{-2} $ as marked in Fig. 4(c). Note that the $ n $ can be on the level of 10$ ^{12} $ cm$^{ -2 }$ in monolayer transition-metal dichalcogenides\cite{Jia, Sun}.

Although no structural distortions have been found in MXenes\cite{Anasori,Meshkian} and the coupling of zero-momentum excitons to the lattice is usually insignificant, we go a step further here and discuss in general terms the consequences of assuming that excitonic instabilities induce structural distortions. Unambiguously, such distortion is a second-order effect occurring on potential excitonic insulators due to energy transfer from excitons to phonons. As the upper limit of this transfer energy is $\mid$$E_t$$\mid$ (The negative of the exciton excitation energy), the back reaction of the lattice distortion would cause a decrease in $\mid$$E_t$$\mid$ but without changing the onset of the excitonic instability. According to the calculations above, the decrease in $\mid$$E_t$$\mid$ leads to a decrease in the $T_{c}$.

Besides the extremely high $ T_{c} $, we find that the exciton BEC is robust against in-plane strain and vertical electric-field. Again, we show for the  Mo$ _{2} $HfC$ _{2} $O$ _{2 } $ the DFT bands under $\pm$1$ \% $ strain, and under 0.2 and 1.2 V/\AA\ electric-field in Figs. 4(d) and 4(e), respectively. It can be seen that the gap feature keeps unchanged for all the cases, and so do the nature of non-trivial topology ($ Z_{2} $ = 1) and the occurrence of excitonic instability ($  E_{t} < $ 0). Quantitatively, the $  E_{t} $ changes very small, within 0.03 eV, as shown in Table S2\cite{SI}. Such an ideal bulk insulation, together with the ultrahigh transition temperature, is important for developing low-dissipative electronic devices and realizing novel topological applications.

From a materials perspective, at least 24 different double transition-metal carbides have been theoretically predicted\cite{ Jin}, of which Mo$ _{2} $TiC$ _{2} $T$ _{x} $, Mo$ _{2} $ScC$ _{2} $T$ _{x} $, Cr$ _{2} $TiC$ _{2} $T$ _{x} $ and Mo$ _{2} $Ti$ _{2} $C$ _{3} $T$ _{x} $ have been synthesized\cite{Anasori,Meshkian} (T denotes different surface termination groups such as O, halogen or OH, and $ x $ is variable). Encouragingly, our calculations show that the Mo$ _{2} $TiC$ _{2} $O$ _{2} $ is a room temperature TEI at the minimum exciton density  $\sim$10$ ^{5} $ cm$ ^{-2} $, which makes the Mo$ _{2} $TiC$ _{2} $T$ _{x} $ perhaps the most promising MXene to experimentally verify our present findings. In fact, there are many other topological insulators with the SOC-independent band inversion, such as transition-metal halides\cite{ Zhou} and functionalized group-IV monolayer\cite{ Xu}, not limited to the MXenes, which are all potential TEIs.

In conclusion, we show that the coexistence of non-trivial topology and excitonic instability in centrosymmetric crystals encounters parity frustration, which poses a hindrance to the TEI formation. Potential TEI candidates should have characteristic one-electron property of ``functional segregation'', e.g., conventional topological insulators with the SOC-independent band inversion. This is confirmed by first-principles $GW$-BSE calculations on the existing MXenes, which are predicted to be intrinsic TEIs with large many-body gap, high transition temperature, and robustness to strain and external electric-field. Also, we give the scheme for experimental confirmation. Once confirmed as a TEI, it will not only solve the problem of room temperature topological insulators, but also realize the first undisputed excitonic insulator, a puzzle pursued for more than half a century. These TEIs can also offer other interesting prospects for both fundamental physics and technological applications due to the macroscopic quantum properties at room temperature.

\begin{acknowledgments}
This work was supported by the Ministry of Science and Technology of China (Grant No. 2020YFA0308800) and the National Natural Science Foundation of China (Grant No. 12074034).
\end{acknowledgments}

\end{document}